\documentclass[letterpaper, 10 pt, conference]{ieeeconf}  

\IEEEoverridecommandlockouts                              

\usepackage{cite}
\usepackage{array} 
\usepackage{amsmath,amssymb,amsfonts}
\usepackage{graphicx}
\let\labelindent\relax
\usepackage{enumitem}
\usepackage{textcomp}
\usepackage{xcolor}

{}
{}
\newtheorem{remark}{Remark}{}

 \usepackage{hyperref}
 \usepackage{booktabs}
 \usepackage{float}
\usepackage{placeins}
\usepackage{float} 
\usepackage{algorithm}
\usepackage{algpseudocode}
\usepackage{array, threeparttable, tabularx, booktabs, caption}
\usepackage{stfloats} 
\title{\LARGE \bf
An HMDP-MPC Decision-making Framework with Adaptive Safety Margins and Hysteresis for Autonomous Driving
}

\author{Siyuan~Li$^{1}$,
        Chengyuan~Liu$^{1}$*,
        and~Wen-hua~Chen$^{2}$
\thanks{$^{1}$ Siyuan Li and Chengyuan Liu are with the Department of Aeronautical and Automotive Engineering, Loughborough University, Loughborough, UK. (e-mail: s.li8@lboro.ac.uk, c.liu10@lboro.ac.uk).}
\thanks{$^{2}$Wen-hua Chen is with the Department of Aeronautical and Aviation Engineering, The Hong Kong Polytechnic University, Hong Kong. ( e-mail: wenhua.chen@polyu.edu.hk).}
}

\begin{document}

\maketitle
\thispagestyle{empty}
\pagestyle{empty}

\begin{abstract}

This paper presents a unified decision-making framework that integrates Hybrid Markov Decision Processes (HMDPs) with Model Predictive Control (MPC), augmented by velocity-dependent safety margins and a prediction-aware hysteresis mechanism. Both the ego and surrounding vehicles are modeled as HMDPs, allowing discrete maneuver transition and kinematic evolution to be jointly considered within the MPC optimization. Safety margins derived from the Intelligent Driver Model (IDM) adapt to traffic context but vary with speed, which can cause oscillatory decisions and velocity fluctuations. To mitigate this, we propose a frozen-release hysteresis mechanism with distinct trigger and release thresholds, effectively enlarging the reaction buffer and suppressing oscillations. Decision continuity is further safeguarded by a two-layer recovery scheme: a global bounded relaxation tied to IDM margins and a deterministic fallback policy. The framework is evaluated through a case study, an ablation against a no-hysteresis baseline, and large-scale randomized experiments across 18 traffic settings. Across 8,050 trials, it achieves a collision rate of only 0.05\%, with 98.77\% of decisions resolved by nominal MPC and minimal reliance on relaxation or fallback. These results demonstrate the robustness and adaptability of the proposed decision-making framework in heterogeneous traffic conditions.

\end{abstract}

\section{INTRODUCTION}

Decision-making in autonomous driving takes place in dynamic and uncertain environments, where the ego vehicle (EV) must generate safe and anticipatory maneuvers while continuously interacting with surrounding vehicles (SVs) \cite{chen2024review,wang2023decision}. A wide spectrum of approaches has been explored, ranging from rule-based heuristics \cite{shu2025decision} to learning-driven policies \cite{li2025investigation} and optimization-based planning \cite{cheng2024hierarchical}, each balancing adaptability, interpretability, and safety in different ways. Among these, Model Predictive Control (MPC) is particularly appealing due to its ability to encode vehicle dynamics and safety constraints within a transparent optimization framework. Existing MPC-based decision-making methods, whether focusing primarily on the EV with limited treatment of surrounding-agent uncertainty \cite{wang2025high,li2024integrated,ammour2022mpc} or incorporating multi-agent predictions into the planning loop \cite{zhang2025mobil,zhang2024interaction,zhou2025robust}, still tend to rely on scenario-dependent heuristics or fixed safety thresholds. 
Such designs often become brittle when deployed across heterogeneous roads layouts, varying traffic densities, and different speed regimes. This motivates the development of a unified MPC framework that adapts safety modeling online and ensures reliable decision-making under diverse traffic conditions.

A key challenge in extending decision-making to complex traffic is determining safe inter-vehicle distances. Fixed headway rules \cite{gilbert2021multi} are widely adopted for simplicity, yet their static nature makes them overly conservative, limiting efficiency and adaptability. Control Barrier Functions (CBFs) \cite{ames2019control,alan2023control,allamaa2024real} provide formal safety guarantees at the low-level control scale, while their effectiveness at high-level planning scale is limited by slow information updates, and the additional variables introduced by the barrier constraints significantly increase solver complexity.
Unlike the methods discussed above, the Intelligent Driver Model (IDM) \cite{treiber2000congested} provides a simple yet effective approach to specify safety margins at the planning layer. IDM generates velocity-dependent distances that adapt to traffic context, making it applicable from free-flowing highways to congested urban roads. However, this approach introduces two challenges. First, the dynamic margins can fluctuate rapidly, leading to frequent threshold crossings that manifest as oscillatory decision switching and velocity variations. Second, the tighter, context-sensitive distances may create infeasible cases in dense or highly uncertain scenarios, where fixed margins would remain feasible but overly conservative. These observations highlight the need for mechanisms beyond IDM to stabilize decisions and ensure feasibility across diverse traffic conditions.

Several lines of research have sought to address the two practical issues described above. To curb decision shifts under time-varying safety thresholds, prior studies smooth speed references or safety margins \cite{liu2024comfort}, impose dwell-time or hysteresis \cite{wei2010prediction}, and shape costs to penalize frequent mode switches and unnecessary lane changes \cite{zheng2024highway}. These tactics can attenuate high-frequency oscillations, but they often rely on high-rate state updates unavailable at planning layer, introduce extra decision variables and with additional solver burden, or remain decoupled from multi-step prediction and multi-agent interactions. For the infeasibility problem, common remedies include soft-constraint penalties \cite{zhao2024real}, constraint tightening or tube-based robust design \cite{samada2023robust}, horizon reduction with warm starts, and predefined fallback maneuvers (e.g., braking or freezing lateral motion). In practice, these fixes are frequently applied in isolation and tuned ad hoc: slacks may be unbounded or weakly tied to the physical meaning of the safety distance, tightening requires disturbance sets that are difficult to calibrate across traffic regimes, and shortening the horizon undermines anticipatory maneuvers. As a result, these patches provide little guidance on when to relax, by how much, and with what effect on safety semantics, and they are rarely stress-tested across heterogeneous scenario families \cite{geurts2023model}. Collectively, these limitations highlight the need for a unified, adaptive framework that can suppress oscillatory switching and ensure robust, continuous operation across diverse traffic conditions.

 To mitigate the oscillation issues described above, we combine IDM-based safety margins with a prediction-aware hysteresis mechanism. The IDM distance provides velocity-dependent margins, and both the trigger and release thresholds are anchored to this baseline, thereby retaining its dependence on speed and traffic context. A trigger threshold activates a corrective regime when the predicted relative gap over the horizon falls below it. Once activated, the regime can be exited only after the gap exceeds a larger release threshold, which effectively enlarges the safety margin, provides the EV with more reaction distance for stable decision-making, and suppresses oscillatory switching and velocity fluctuations. To address potential infeasibility, decision continuity is further maintained outside the nominal MPC loop through a two-layer recovery scheme: a global bounded relaxation tied to the IDM margin and, if necessary, a deterministic rule-based fallback.

Extending our previous studies, both the EV and SVs are modeled as HMDPs to capture discrete maneuver transitions and kinematic evolution, which are then embedded within the MPC framework. On this unified basis, the safety margins, hysteresis logic, and recovery scheme are co-designed, ensuring consistent operation as an integrated MPC framework rather than a collection of isolated fixes. This holistic integration enhances adaptability to diverse traffic conditions and improves robustness for practical deployment. The effectiveness of the framework is demonstrated through extensive case studies and large-scale randomized experiments.

The main contributions of this paper are as follows:
\begin{itemize}

\item An integrated decision-making framework is established, where the EV and SVs are jointly modeled as HMDPs, and the resulting multi-modal predictions are seamlessly embedded into an MPC formulation for planning under uncertainty.
    \item IDM-based safety margins are combined with a frozen-release hysteresis mechanism, which consistently enlarges the effective reaction distance and suppresses oscillatory decision switching and velocity fluctuations.
    \item A two-layer recovery scheme is proposed, consisting of a global bounded relaxation and a deterministic rule-based fallback, ensuring continuity of decisions.
    \item A comprehensive evaluation is conducted, including representative case studies, ablation experiment, and large-scale randomized testing, to demonstrate robustness and practicality under heterogeneous traffic conditions.
\end{itemize}

The remainder of the paper is organized as follows. Section II introduces the overall system architecture. Section III presents the HMDP modeling, safety constraints, the hysteresis mechanism, the MPC formulation with cost function design, and recovery scheme. Section IV reports case studies, ablation experiment, and large-scale randomized evaluations. Section V concludes the paper.


\textbf{Notation.} For an integer $n \ge 1$, let 
$\mathcal{I}_n := \{1,2,\dots,n\}$, where $n$ denotes the total number of lanes. Define the set $\mathcal{M} := \{-1,0,1\}$, 
where the values $-1$, $0$, and $1$ are interpreted according to context. 
Its Cartesian square is 
$\mathcal{M}^2 := \{(a,b)\mid a\in\mathcal{M},\; b\in\mathcal{M}\}$. The indicator function $\mathbb{I}(\cdot)$ equals $1$ if the condition inside holds true, and $0$ otherwise.

\section{System Overview}

\begin{figure}[t]
    \centering
    \includegraphics[width=\linewidth]{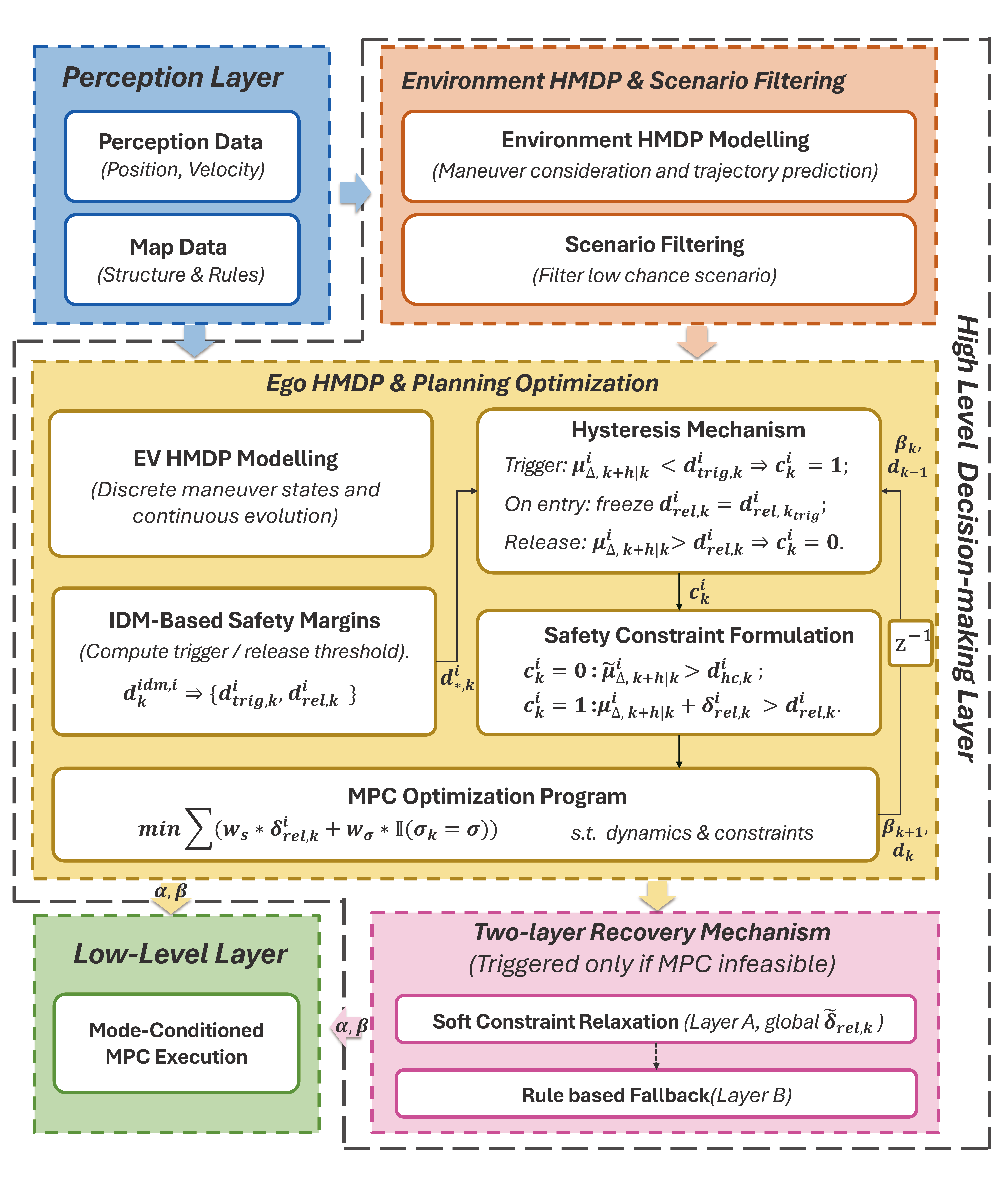}
    \caption{Unified prediction-aware decision-making framework.
    Formal definitions of symbols are provided in subsequent sections.}
    \label{fig:framework}
\end{figure}

The overall architecture of the proposed  decision-making framework is shown in Fig.~\ref{fig:framework}.
Perception outputs serve two key functions in the framework. First, they provide the state information to the HMDP-based environment model, which estimates SV maneuvers and predicts their future trajectories, with low-probability outcomes pruned to reduce computational complexity. Second, the same perception information is used by the IDM to compute velocity-dependent safety margins for the EV. To prevent these dynamic margins from inducing oscillatory decision switching, a prediction-aware hysteresis mechanism is incorporated into the safety constraint formulation. Based on the predicted environment information and safety constraints, the MPC optimization is solved online to produce high-level decisions for the EV. When the nominal optimization becomes infeasible, decision continuity is preserved through a two-layer recovery scheme: a global bounded relaxation tied to the IDM margin, followed by a deterministic rule-based fallback. Finally, the resulting high-level decisions are executed by the low-level control modules.

\section{Problem Formulation}
\subsection{HMDP Modeling of EV}
In this work, the EV’s high-level maneuver is represented as the state of an MDP within the HMDP framework. At each time step $k$, the state is given by
\begin{equation*}
m_k^{\text{ev}}=(\alpha_k^{\text{ev}},\beta_k^{\text{\,ev}}),
\end{equation*}
where $\alpha_k^{\text{ev}} \in \mathcal{I}_n$ denotes the target lane index occupied by the EV after the maneuver, 
$\beta_k^{\text{\,ev}} \in \mathcal{M}$ represents the longitudinal maneuver state, where the elements $-1$, $0$, and $1$ correspond to decelerating, cruising (constant speed), and accelerating, respectively.


At each time step, the EV selects a discrete action
\begin{equation}\label{eq:action}
d_k^{\text{\,ev}}
=(d_k^{\text{\,ev,lat}},\,d_k^{\text{\,ev,long}})\in\mathcal{M}^2,
\end{equation}
where $d_k^{\text{\,ev,lat}}\in\mathcal{M}$ denotes the action in lateral, with the elements $-1$, $0$, and $1$ correspond to left lane change, lane keeping, and right lane change, respectively. The
$d_k^{\text{\,ev,long}}\in\mathcal{M}$ denotes the action in longitudinal, with the elements $-1$, $0$, and $1$ correspond to decelerate, keep constant-speed, and accelerate, respectively. 
This action governs the transitions between maneuver states, subject to the restriction that only one-step transitions are allowed, thereby preventing abrupt simultaneous changes in both lateral and longitudinal directions.

The discrete transitions between maneuver states are formulated as
\begin{equation}\label{eq:state_transition}
\alpha_{k+1}^{\text{ev}}=\operatorname{clip}(\alpha_k^{\text{ev}}+d_k^{\text{\,ev,lat}}),~~
\beta_{k+1}^{\text{\,ev}}=\operatorname{clip}(\beta_k^{\text{\,ev}}+d_k^{\text{\,ev,long}}),
\end{equation}
where the operator $\operatorname{clip}(\cdot)$ prunes infeasible actions, if the 
resulting value lies outside the admissible set, the corresponding action is discarded. 
This guarantees that $\alpha_{k+1}^{\mathrm{ev}}\in\mathcal I_n$ and 
$\beta_{k+1}^{\mathrm{ev}}\in\mathcal M$ at every step $k$.

The discrete maneuver transitions not only update the EV’s maneuver state but also determine the kinematic map for propagating its kinematic states.
At each discrete step $k$, we denote the EV’s kinematic states as
\begin{equation*}
\mathbf{x}_k^{\text{ev}}=[\,x_k^{\text{ev}},\,y_k^{\text{ev}},\,v_k^{\text{ev}}\,]^\top
\end{equation*}
 where $x_k^{\text{ev}}$ and $y_k^{\text{ev}}$ denote the the longitudinal and lateral positions, and $v_k^{\text{ev}}$ is the longitudinal velocity. We adopt a hybrid selector-indexed family of discrete-time kinematic maps
\[
\mathcal{F}_{\text{kin}}=\{\,f_{\text{kin}}^{\sigma}:\mathbb{R}^3\!\to\!\mathbb{R}^3 \,\mid\, \sigma\in\mathcal{M}^2\},
\]
and at step $k$ the selector
\(
\sigma_k=(d_k^{\text{\,ev,lat}},\,\beta_{k+1}^{\text{\,ev}})
\)
chooses the active map, where the lateral action and longitudinal
maneuver state are combined asymmetrically; see Remark~\ref{remark} (Section.~\ref{sectionE}) for details. The kinematic states are updated by
\begin{equation}\label{eq:ev_dynamics}
\mathbf{x}_{k+1|k}^{\text{ev}}=f_{\text{kin}}^{\sigma_k}(\mathbf{x}_k^{\text{ev}},u_k^{\text{ev}}),
\end{equation}
i.e., different $(d_k^{\text{\,ev,lat}},\beta_{k+1}^{\text{\,ev}})$ pairs correspond to different kinematic maps, while $u_k^{\text{ev}}$ denotes the associated kinematic input, together yielding distinct trajectory predictions.

\subsection{HMDP Modeling of SVs with Scenario Pruning}

Each SV’s high-level maneuver is also represented as the state of an MDP within the HMDP framework. At each time step $k$, the discrete maneuver state is
\begin{equation*}
m_k^{\text{sv}} = (\alpha_k^{\text{sv}}, \beta_k^{\text{\,sv}}),
\end{equation*}
with $\alpha_k^{\text{sv}} \in \mathcal{I}_n$ denoting the lateral maneuver state and 
$\beta_k^{\text{\,sv}} \in \mathcal{M}$ denoting the longitudinal maneuver state. 
The corresponding action is
\begin{equation}\label{eq:svaction}
d_k^{\text{\,sv}}=(d_k^{\text{\,sv,lat}},\,d_k^{\text{\,sv,long}})\in\mathcal{M}^2,
\end{equation}
and the maneuver transition is given by
\begin{equation}\label{eq:svtransition}
m_{k+1}^{\text{sv}}
= \Big(\operatorname{clip}(\alpha_k^{\text{sv}}+d_k^{\text{\,sv,lat}}),\;
       \operatorname{clip}(\beta_k^{\text{\,sv}}+d_k^{\text{\,sv,long}})\Big).
\end{equation}

The kinematic state of the $i$-th SV is defined consistently as
\begin{equation*}
\mathbf{x}_k^{\text{sv},i}=[\,x_k^{\text{sv},i},\,y_k^{\text{sv},i},\,v_k^{\text{sv},i}\,]^\top,
\end{equation*}
representing its
position and longitudinal speed. 
The kinematic states are updated by
\begin{equation}\label{eq:svcontinuous}
\mathbf{x}_{k+1|k}^{\text{sv},i}
= f_{\text{kin}}^{\sigma_k^{\text{sv},i}}\!\big(\mathbf{x}_k^{\text{sv},i},u_k^{\text{sv},i}\big) + w_k^i,
\end{equation}
where $\sigma_k^{\text{sv},i}=(d_k^{\text{\,sv,lat}},\,\beta_{k+1}^{\text{sv}})$ indexes the kinematic map, 
$u_k^{\text{sv},i}$ denotes the nominal kinematic input corresponding to the assumed maneuver, and $w_k^i \sim \mathcal{N}(0,\Sigma_k^i)$ denotes the modeling uncertainty, where $\Sigma_k^i$ is the covariance matrix.

At each step, multiple candidate trajectories are generated for each SV by rolling out \eqref{eq:svcontinuous} under different maneuver hypotheses. Low-probability branches are pruned, and the remaining trajectories define the interaction space used to assess the safety of the EV’s decisions.

\subsection{Safety Constraints}

Safety constraints are imposed to ensure that the EV maintains a safe distance from SVs whenever they are predicted to occupy the same lane. 
Unlike fixed-distance margins, we adopt velocity-dependent thresholds and a probabilistic formulation to adapt to diverse traffic conditions. 

Specifically, the minimum safe gap relative to the $i$-th SV is computed using the IDM:
\begin{equation*}\label{idm_gap}
d_k^{\,\text{idm},i} = d_0 + v_k^{\text{ev}} T + \frac{v_k^{\text{ev}} \Delta v_k^i}{2\sqrt{ab}},
\end{equation*}
where $d_0$ is the jam spacing, $T$ is the desired headway, $a$ is the maximum acceleration, $b$ is the comfortable deceleration, and 
$\Delta v_k^i = v_k^{\text{ev}} - \bar{v}_k^{\text{sv},i}$ is the relative velocity with $\bar{v}_k^{\text{sv},i}$ denoting the nominal velocity of the SV, which corresponds to the mean prediction without the additive noise.

With additive Gaussian noise in \eqref{eq:svcontinuous} ($w_k^i\!\sim\!\mathcal N(0,\Sigma_k^i)$), the one-step SV prediction is Gaussian, whereas the EV state is treated as deterministic under MPC prediction. 
Consequently, the relative longitudinal distance
\[
\Delta x_{k+1|k}^i = x_{k+1|k}^{\text{sv},i}-x_{k+1|k}^{\text{ev}} 
\]
is also Gaussian with mean
\[
\mu_{\Delta,k+1|k} ^{\,i}= \bar{x}_{k+1|k}^{\text{sv},i}-x_{k+1|k}^{\text{ev}}
\]
and variance $\Sigma_{\Delta,k+1|k}$, corresponding to the longitudinal component of $\Sigma_{k+1|k}^i$,
where $\bar{x}_{k+1|k}^{\text{sv},i}$ denotes the predicted mean longitudinal position of the $i$-th SV.

To handle both cases (EV ahead versus EV behind), we introduce a sign variable 
$s_k^i \in \{+1,-1\}$, where $s_k^i=-1$ if the EV is ahead of the $i$-th SV and 
$s_k^i=+1$ if the EV is behind of the $i$-th SV. We then define
\[
\tilde{\Delta}x_{k+1|k}^i: = s_k^i\,\Delta x_{k+1|k}^i,
\]
\begin{equation}\label{eq:tildemu}
    \tilde{\mu}_{\Delta,k+1|k}^{\,i}: = s_k^i\,\mu_{\Delta,k+1|k} ^{\,i}.
\end{equation}

Based on the signed gap defined above, the corresponding safety chance constraint can be formulated as
\begin{equation*}\label{eq:chance}
\mathbb{P}\!\left(\tilde{\Delta}x_{k+1|k}^i \geq d_k^{\,\text{idm},i}\right) \geq 1-\epsilon,
\end{equation*}
since $\tilde{\Delta}x_{k+1|k}^i$ is Gaussian (a signed linear transformation of $\Delta x_{k+1|k}^i$), the chance constraint can be reformulated in deterministic form (see e.g. \cite{Calafiore2006DRCC}) as
\begin{equation}\label{eq:deterministic}
\tilde{\mu}_{\Delta,k+1|k}^{\,i} \;\geq\; d_k^{\,\text{idm},i} + z_\epsilon \sqrt{\Sigma_{\Delta,k+1|k}},
\end{equation}
where $z_\epsilon$ is the standard-normal quantile corresponding to confidence $1-\epsilon$.


\subsection{Hysteresis Mechanism}

While the IDM-based safety distance $d_k^{\,\text{idm},i}$ enhances adaptability across diverse traffic scenarios, its direct dependence on ego velocity can cause frequent decision switching. For example, if the relative distance falls slightly below the safety threshold, the EV is forced to brake. The consequent reduction in velocity decreases $d_k^{\,\text{idm},i}$ in the following step, which may prompt a switch back to cruising or acceleration. This back-and-forth maneuver results in oscillations in both maneuver choices and velocity profiles. To suppress such oscillations, we employ a hysteresis mechanism, if any predicted gap falls below a trigger threshold, the system enters a corrective regime that requires a larger inter-vehicle distance, it leaves only when the gap exceeds a higher release threshold. 


First, a time-varying distance bandwidth is computed as
\begin{equation*}
\varepsilon_{\text{trig},k}^{i} = \min \!\Big( \max(k_\varepsilon d_k^{\,\text{idm},i},\, \varepsilon_{\min}),\, \varepsilon_{\max} \Big),
\end{equation*}
where $\varepsilon_{\text{trig},k}^{i}$ represents an adaptive distance margin around the hard safety distance $d_k^{\,\text{idm},i}$. 
Here, $k_\varepsilon \in (0,1)$ sets a relative margin proportional to $d_k^{\,\text{idm},i}$, while $\varepsilon_{\min}$ and $\varepsilon_{\max}$ are absolute bounds to avoid overly small/large margins.
The corresponding trigger and release distance offsets are defined as
\begin{equation*}
\Delta_{\text{trig},k}^{i} = \gamma_1 \varepsilon_{\text{trig},k}^{i}, 
\qquad 
\Delta_{\text{rel},k}^{i} = \gamma_2 \varepsilon_{\text{trig},k}^{i},
\end{equation*}
where $\gamma_1<\gamma_2$ yields a wider release band than the trigger band, preventing Zeno-like switching \cite{Berkane2021Zeno}.

The corresponding trigger and release distances are then expressed as
\begin{equation*}
d_{\text{trig},k}^{\,i} = d_{\mathrm{hc},k}^{\,i} + \Delta_{\text{trig},k}^{i}, 
\qquad 
d_{\text{rel},k}^{\,i} = d_{\mathrm{hc},k}^{\,i} + \Delta_{\text{rel},k}^{i},
\end{equation*}
where
\begin{equation*}
d_{\mathrm{hc},k}^{\,i} = d_k^{\,\text{idm},i} + z_\epsilon \sqrt{\Sigma_{\Delta,k+1|k}}.
\end{equation*}
Here, $d_{\mathrm{hc},k}^{\,i}$ denotes the deterministic safety margin, which serves as the reference for constructing the hysteresis bands.

We then introduce a binary variable $c_k^{i} \in \{0,1\}$ to indicate whether the EV is in the corrective regime with respect to the $i$-th SV at time step $k$.  
The switching logic is determined by evaluating the predicted longitudinal relative distances $\mu_{\Delta,k+h|k}^i$ over the MPC horizon $\mathcal{H}=\{1,\dots,H\}$:

\begin{equation*}\label{eq:corrective_mode}
c_k^{i} =
\begin{cases}
1, 
& \text{if } \exists h\in\mathcal{H}:\ 
\mu_{\Delta,k+h|k}^{i} < d_{\text{trig},k}^{\,i}, \\[3pt]
0, 
& \text{if } c_{k-1}^{i}=1 \land 
\exists h\in\mathcal{H}:\ 
\mu_{\Delta,k+h|k}^{i} \ge d_{\text{rel},k}^{\,i}, \\[3pt]
c_{k-1}^{i}, 
& \text{otherwise.}
\end{cases}
\end{equation*}
Here the trigger threshold $d_{\text{trig},k}^{\,i}$ activates the corrective mode when a predicted violation occurs,  
while the release threshold $d_{\text{rel},k}^{\,i}$ defines the condition for exiting the corrective mode once the predicted distance is sufficiently restored. 

However, since the release threshold $d_{\text{rel},k}^{\,i}$ depends on the ego velocity, it may shrink as the EV decelerates. 
This can lead to premature exit from the corrective regime and repeated reentry, causing undesirable oscillations. 
To prevent this,  the release threshold is frozen at the moment when the corrective mode is first activated. 
Let $k_{\text{trig}}^{i}$ denote the triggering time for the $i$-th SV, and define
\begin{equation*}
d_{\text{rel}}^{\,i} := d^{\,i}_{\text{rel},k_{\text{trig}}}.
\end{equation*}
By fixing the release threshold in this way, the corrective mode is maintained, when the EV is following the $i$-th SV, until the EV reaches a genuinely safe distance, i.e.,
\begin{equation*}\label{eq:frozen_release}
\mu_{\Delta,k+h|k}^{\,i} \;\ge\; d^{\,i}_{\text{rel},k_{\text{trig}}}, 
\qquad \forall\, h \in \mathcal{H}.
\end{equation*}
Here, $d^{\,i}_{\text{rel},k_{\text{trig}}}$ is treated as a stabilizing buffer to prevent oscillatory switching in corrective mode, while safety is ensured by the deterministic constraint in~\eqref{eq:deterministic}. Its role is to provide an additional reaction margin beyond the deterministic safety distance, therefore, strict enforcement is unnecessary. Based on this observation, we relax the tightened constraint by introducing a slack variable $\delta_{\text{rel},k}^{\,i}$, which is active only in the corrective mode and bounded by the hysteresis bandwidth:
\begin{equation*}
\delta_{\text{rel},k}^{\,i} \in
\begin{cases}
[0,\; d_{\mathrm{HS}}^{\,i}], & \text{if } c_k^{\,i}=1,\\
0, & \text{if } c_k^{\,i}=0,
\end{cases}
\quad
d_{\mathrm{HS}}^{\,i} \;:=\; d^{\,i}_{\text{rel},k_{\text{trig}}} - d^{\,i}_{\text{trig},k_{\text{trig}}}.
\end{equation*}

The relaxed safety constraint is then formulated as
\begin{equation}\label{eq:soft_safety}
\mu_{\Delta,k+h|k}^{\,i} + \delta_{\text{rel},k}^{\,i}
\;\;\geq\;\; d^{\,i}_{\text{rel},k_{\text{trig}}}, 
\qquad \forall h \in \mathcal{I}_H,
\end{equation}
which applies only when the EV is behind the $i$-th SV and the corrective mode is active 
(i.e., $c_k^{\,i}=1 \wedge s_k^i=1$). In contrast, when corrective mode is inactive ($c_k^{\,i}=0$) or the EV is ahead of the $i$-th SV ($s_k^i=-1$), 
the standard chance-constrained form in~(\ref{eq:deterministic}) is adopted instead.

Finally, the slack variable $\delta_{\text{rel},k}^{\,i}$ is penalized in the MPC cost function, ensuring that relaxation serves as a controlled buffer around the release margin.

\subsection{MPC Formulation with Cost Function Design}\label{sectionE}

Building on the hysteresis mechanism for safety constraints, we now define the MPC objective that (i) penalizes the hysteresis slack to eliminate violations of the frozen-release threshold and facilitate exit from the corrective regime, and (ii) represents maneuver preferences directly through the discrete hybrid selector pair \(
\sigma_k=(d_k^{\text{\,ev,lat}},\,\beta_{k+1}^{\text{\,ev}})
\).

The hysteresis slack \(\delta_{\text{rel},k}^{\,i}\) only activate in corrective mode and bounded by the bandwidth \(d_{\text{HS}}^{\,i}\), and contributes to the cost by
\begin{equation*}
J^{\text{safety}}
=\sum_{k=0}^{H-1}\sum_{i=1}^{N_{\text{sv}}}
w_s\,\delta_{\text{rel},k}^{\,i}.
\end{equation*}
This penalization drives the predicted gap to exceed the release threshold as early as feasible.

To capture high-level maneuver preferences, each \(\sigma_k\) is assigned a weight \(w_\sigma\). The resulting maneuver term is
\begin{equation*}
J^{\text{beh}}
=\sum_{k=0}^{H-1}\sum_{\sigma\in\mathcal{M}^2}
w_\sigma\,\mathbb{I}(\sigma_k=\sigma),
\end{equation*}
where \(\mathbb{I}(\cdot)\) is the indicator function. The weights $w_\sigma$ is designed with a safety-first, efficiency-second principle: conservative combinations, such as (keep lane, cruising), are assigned lower weights, while aggressive ones, such as (accelerate, lane change), incur higher penalties.  

Based on the objective terms and all constraints on maneuvers transition, kinematic evolution, and safety, the complete MPC optimization program is formulated as
\begin{equation}
\begin{aligned}
\min_{d_k^{\,\mathrm{ev}},\,m_k^{\mathrm{ev}},\,\delta^{\,i}_{\mathrm{rel},k}}\quad
& \sum_{k=0}^{H-1}\!\left(\sum_{i=1}^{N_{\mathrm{sv}}} w_s\,\delta^{\,i}_{\mathrm{rel},k}
   + \sum_{\sigma} w_\sigma\,\mathbb{I}(\sigma_k=\sigma)\right) \\[0.2em]
\text{s.t.}\quad
& \alpha_k^{\mathrm{ev}} \in \mathcal{I}_n,\quad
  \beta_k^{\mathrm{ev}} \in \mathcal{M}, \\ \label{eq:evstate}
&
  \alpha_k^{\mathrm{sv}} \in \mathcal{I}_n,\quad
  \beta_k^{\mathrm{sv}} \in \mathcal{M}, \\
&
  \eqref{eq:action}\text{--}\eqref{eq:soft_safety}.
\end{aligned}
\end{equation}

\begin{remark}\label{remark}
The hybrid selector \(\sigma_k=(d_k^{\text{\,ev,lat}},\,\beta_{k+1}^{\text{ev}})\) deliberately combines a lateral action with a longitudinal state. 
For lateral decisions, no inherent preference is assigned to specific lanes; the key consideration is whether a lane change occurs. Accordingly, lane-change actions are penalized directly to discourage unnecessary maneuvers.
In contrast, longitudinal preferences are better represented as maneuver states, since the design goal is to promote steady cruising and minimize frequent throttle or brake switches. 
This asymmetric design reflects the semantics of driving: lateral choices are action-driven, whereas longitudinal evolution is maneuver-driven.
\end{remark}

\subsection{Two-layer recovery scheme}
While the proposed formulation integrates hysteresis-based slack and maneuver preferences to maintain feasibility in most situations, practical deployment may still encounter solver infeasibility. Such infeasibility may stem from multiple sources, including prediction errors under uncertainty and mismatches between high-level prediction model and low-level vehicle dynamics. To handle rare cases where the nominal MPC becomes infeasible, we introduce a two-layer recovery scheme that maintains conservative safety margins and guarantees decision continuity by always yielding a valid action online.

The first layer introduces a global relaxation slack \( \tilde{\delta}_{\text{rel},k}^{\,i}\). Using the signed gap variable $\tilde{\mu}_{\Delta,k+h|k}^i$ defined in~(\ref{eq:tildemu}) and the hysteresis slack $\delta_{\text{rel},k}^i$ specified in~(\ref{eq:soft_safety}), the relaxed safety constraint can be written in the unified form
\begin{equation}
\tilde{\mu}_{\Delta,k+h|k}^{\,i} \;\ge\;
\begin{cases}
d_{\mathrm{hc},k}^{\,i} - \tilde{\delta}_{\text{rel},k}^{\,i}, 
& \!\!\!\!\text{ if $s_k^i=-1$}, \\[6pt]
d_{\text{rel},k}^{\,i} - \tilde{\delta}_{\text{rel},k}^{\,i} - c_k^i \delta_{\text{rel},k}^{\,i}, 
& \!\!\!\!\text{ if $s_k^i= 1$}.
\end{cases}
\label{eq:relaxed_constraint_case}
\end{equation}
The global slack is constrained as
\begin{equation}
0 \;\le\; \tilde{\delta}_{\text{rel},k}^{\,i} \;\le\; \gamma\, d_k^{\text{\,idm},i}, 
\quad \gamma \in (0,1),
\label{eq:global_slack_bound}
\end{equation}
ensuring that the relaxation is scaled relative to the IDM-based safe distance. In this way, the relaxation is automatically normalized by speed and traffic conditions, preventing both excessive relaxation in dense traffic and overly loose bounds in free-flow scenarios.
The corresponding cost term is written as
\begin{equation*}
J^{\text{global}} = \sum_{k=0}^{H-1}\sum_{i=1}^{N_{\text{sv}}} w_q\,\tilde{\delta}_{\text{rel},k}^{\,i}, 
\quad w_q \gg w_s,
\label{eq:global_cost}
\end{equation*}
and the overall objective is updated as
\begin{equation}
J = J^{\text{safety}} + J^{\text{beh}} + J^{\text{global}}.
\label{eq:total_cost}
\end{equation}

\begin{algorithm}[t]
\caption{}
\label{alg:mpc_decision}
\textbf{Input:}  $\mathbf{x}_k^{\text{ev}}$, $m_k^{\text{ev}}$, $\{\mathbf{x}_k^{\text{sv},i}\}$ with lane association, horizon $H$ \\
\textbf{Output:} $d_k^{\text{\,ev}}$, $m_{k+1}^{\text{ev}}$
\begin{algorithmic}[1]
\State \textbf{Filter SVs:} construct active SV set within perception range (front/rear windows).
\State \textbf{Hysteresis refresh:} compute IDM distances $d_k^{\,\text{idm},i}$ and update corrective indicators $c_k^i$ with trigger $d_{\text{trig},k}^{\,i}$ and release $d_{\text{rel},k}^{\,i}$ thresholds.
\State \textbf{MDP constraints:} enforce EV/SV discrete state-action encoding~(\ref{eq:action},\ref{eq:svaction}), transition rules~(\ref{eq:state_transition},\ref{eq:svtransition}), and kinematic evolution~(\ref{eq:ev_dynamics},\ref{eq:svcontinuous}).
\State \textbf{Scenario tree:} enumerate pruned SV action sequences; predict $(\mathbf{x}^{\text{sv},i}_{k+h|k},\Sigma^{\text{sv},i}_{k+h|k})$.
\State \textbf{Safety constraints:} impose the chance-constrained form via its deterministic equivalent (\ref{eq:deterministic}) 
when $c_k^i=0$, and the relaxed form (\ref{eq:soft_safety}) when $c_k^i=1$.
\State \textbf{Objective:} minimize $J^{\text{safety}}{+}J^{\text{beh}}$; solve the mixed-integer linear program (MILP).
\If{feasible}
  \State apply first-step $d_k^{\text{\,ev}}$; set $m_{k+1}^{\text{ev}}$; 
\Else
  \State \textbf{Layer A (global soft):} replace safety by (\ref{eq:relaxed_constraint_case}) with bound (\ref{eq:global_slack_bound}); augment cost by (\ref{eq:total_cost}); re-solve.
  \If{feasible}
    \State apply $d_k^{\text{\,ev}}$; set $m_{k+1}^{\text{ev}}$; 
  \Else
\State \textbf{Layer B (fallback):} apply fallback (\ref{eq:fallback_policy}); 
  \EndIf
\EndIf
\end{algorithmic}
\end{algorithm}
\begin{remark}
Compared with the local hysteresis slack variable $\delta_{\text{rel},k}^i$, which is mode-conditioned (active only in corrective mode), and bounded by $d_{\text{HS}}^{\,i}$, the global slack variable $\tilde{\delta}_{\text{rel},k}^i$ provides a uniformly bounded feasibility buffer independent of corrective mode. Its weight $w_q$ is chosen sufficiently large relative to $w_s$ to discourage unnecessary use of global slack.
\end{remark}

The second layer of the recovery scheme provides a deterministic rule-based fallback when the optimization problem remains infeasible even under global relaxation. The fallback maneuver is designed to be simple and deterministic: the EV selects a conservative longitudinal action depending on whether the most critical threat arises from a front, rear, or lateral vehicle.

Formally, the fallback policy is expressed as
\begin{equation}
\big(d_{k+1}^{\text{\,ev,lat}},\, \beta_{k+1}^{\text{ev}}\big) =
\begin{cases}
(d_k^{\text{\,ev,lat}}, -1), 
& \text{front vehicle braking}, \\[3pt]
(d_k^{\text{\,ev,lat}}, +1), 
& \text{rear vehicle closing}, \\[3pt]
(d_k^{\text{\,ev,lat}}, 0),  
& \text{lateral risk dominates}.
\end{cases}
\label{eq:fallback_policy}
\end{equation}

This fallback  ensures  spacing is increased in the direction of the most critical threat. Together with the global slack, the two-layer scheme preserves feasibility and continuity of the decision framework.
Algorithm~\ref{alg:mpc_decision} presents the integrated HMDP-MPC decision-making procedure with the two-layer feasibility-recovery mechanism.

\section{Simulation}

The effectiveness of the proposed framework is evaluated through three complementary studies: (i) a representative case study, (ii) an ablation analysis quantifying the contribution of the prediction-aware hysteresis, and (iii) large-scale randomized experiments validating robustness under heterogeneous traffic conditions.

\subsection{Case 1: Demonstration in a Specific Scenario}
In this case study, we consider a two-lane traffic environment with five SVs. 
The initial states of the EV and all SVs are summarized in Table~\ref{tab:initial_conditions}, 
while the simulation parameters are listed in Table~\ref{tab:case1_parameters}.
\begin{table}[h]
\centering
\captionsetup{font=footnotesize, skip=6pt}
\caption{Initial conditions (Case~1)}
\setlength{\tabcolsep}{4pt}
\label{tab:initial_conditions}
\begin{threeparttable}
\begin{tabular}{lllll}
\toprule
\textbf{Vehicle} & \textbf{Initial Lane} & $x_0$ (m) & $y_0$ (m) & $v_0$ (m/s) \\ \midrule
EV (Red)        & Lane 2 & 110.77 & 0 & 16.15 \\
SV1 (Blue)       & Lane 1 & 277.93 & 4 & 22.93 \\
SV2 (Green)      & Lane 2 & 175.50 & 0 & 24.54 \\
SV3 (Amber)      & Lane 1 &  74.75 & 4 & 27.24 \\
SV4 (Purple)     & Lane 1 & 383.90 & 4 & 23.51 \\
SV5 (Orange-red) & Lane 2 & 300.51 & 0 & 21.38 \\
\bottomrule
\end{tabular}
\end{threeparttable}
\end{table}
\begin{table}[t]
\centering
\captionsetup{font=footnotesize, skip=2pt}
\caption{Simulation and vehicle parameters (Case~1)}
\setlength{\tabcolsep}{6pt}
\label{tab:case1_parameters}
\begin{threeparttable}
\begin{tabular}{lll}
\toprule
\textbf{Parameter} & \textbf{Description} & \textbf{Value} \\ \midrule
$T_{\mathrm{sim}}$  & Simulation duration                       & 40 s \\
$T_l$               & Low-level control sampling period         & 0.1 s \\
$T_h$               & High-level decision update period         & 0.4 s \\
$H$                 & High-level prediction horizon             & 3 steps (1.2 s) \\
$w_{\mathrm{lane}}$ & Lane width                                & 4 m \\
$\gamma$            & Soft constraint relaxation ratio          & 0.1 \\
$\gamma_1$          & Trigger margin scaling coefficient        & 1 \\
$\gamma_2$          & Release margin scaling coefficient        & 1.4 \\
$ \varepsilon_{\min}$          & Minimum distance margin         & 6 m \\
$ \varepsilon_{\max}$          & Maximum distance margin         & 22 m\\
$w_s$               & Penalty for safety violation              & 100 \\
$w_q$               & Weight for relaxed constraints            & $10^{4}$ \\
\bottomrule
\end{tabular}
\end{threeparttable}
 \vspace{-6pt}
\end{table}

\begin{figure}[!t]
    \centering
 \includegraphics[width=0.9\linewidth]{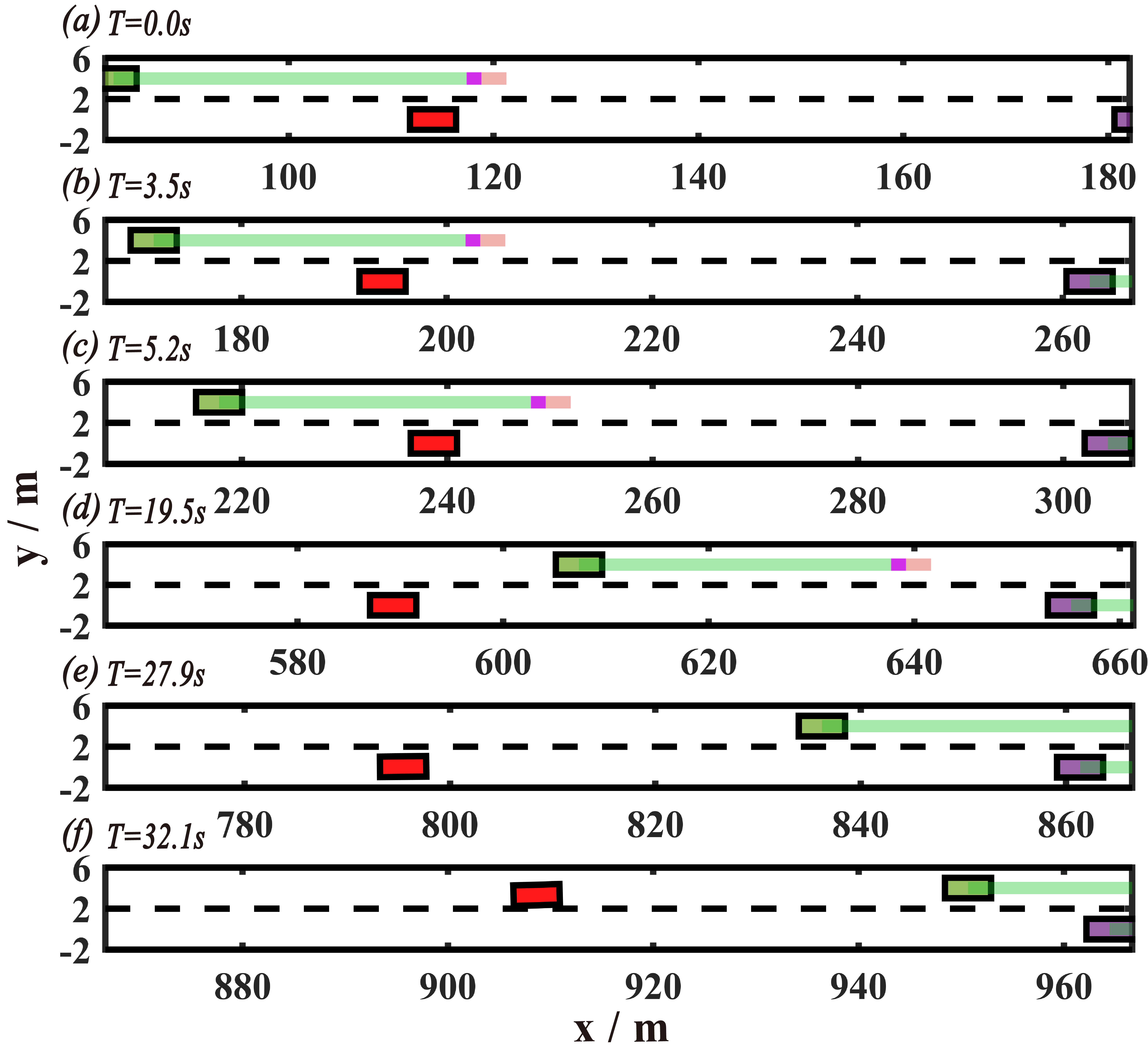}
    \caption{Spatial evolution at representative instants (Case~1). 
Semi-transparent strips indicate predicted longitudinal maneuvers (green: decel; purple: const-speed; red: accel).}
    \label{fig:case1_position}
 \vspace{-6pt}
\end{figure}

\begin{figure}[!t]
    \centering
\includegraphics[width=0.9\linewidth]{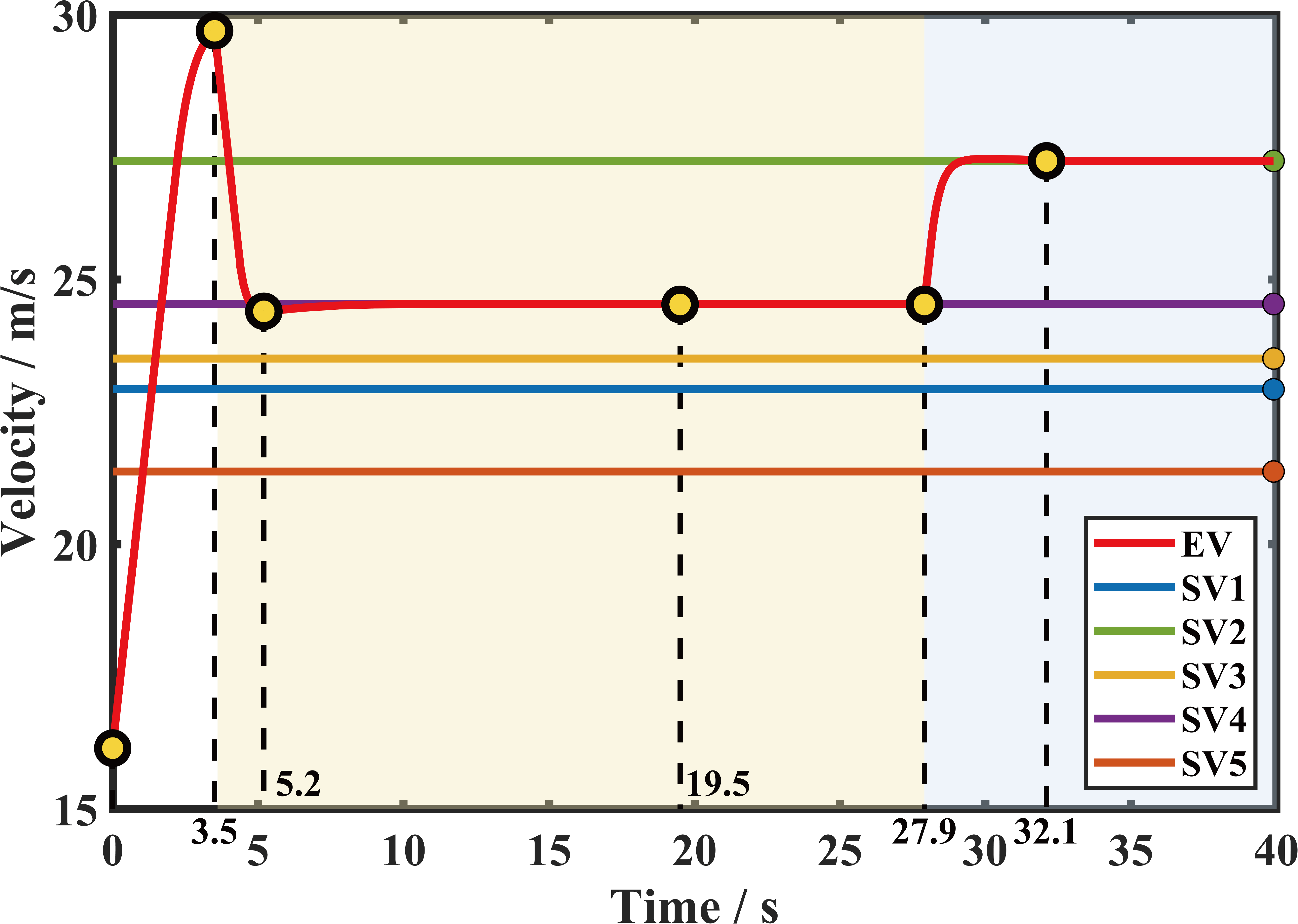}
    \caption{Velocity comparison (EV versus SVs) (Case~1). Yellow circles align with the snapshots in Fig.~\ref{fig:case1_position}.}
    \label{fig:case1_velocity}
    \vspace{-12pt}
\end{figure}

Under the initial conditions, the decision process unfolds in three phases (see Fig.~\ref{fig:case1_position}--\ref{fig:case1_velocity}): 
(i) 0-3.5 s: the EV accelerates smoothly to join traffic; 
(ii) 3.5-27.9 s: after closing in on a slower leader, the trigger threshold is crossed and the corrective mode becomes active (yellow background in Fig.~\ref{fig:case1_velocity}), producing stable gap regulation without oscillatory decision switching; 
(iii) \(\geq\) 27.9 s: a single lane change is executed, the corrective regime remains active but switches to a new reference SV in the target lane (blue background  in Fig.~\ref{fig:case1_velocity}), and the speed settles smoothly.

\subsection{Case 2: Ablation on Hysteresis}

\begin{figure}[t]
    \centering
    \includegraphics[width=0.9\linewidth]{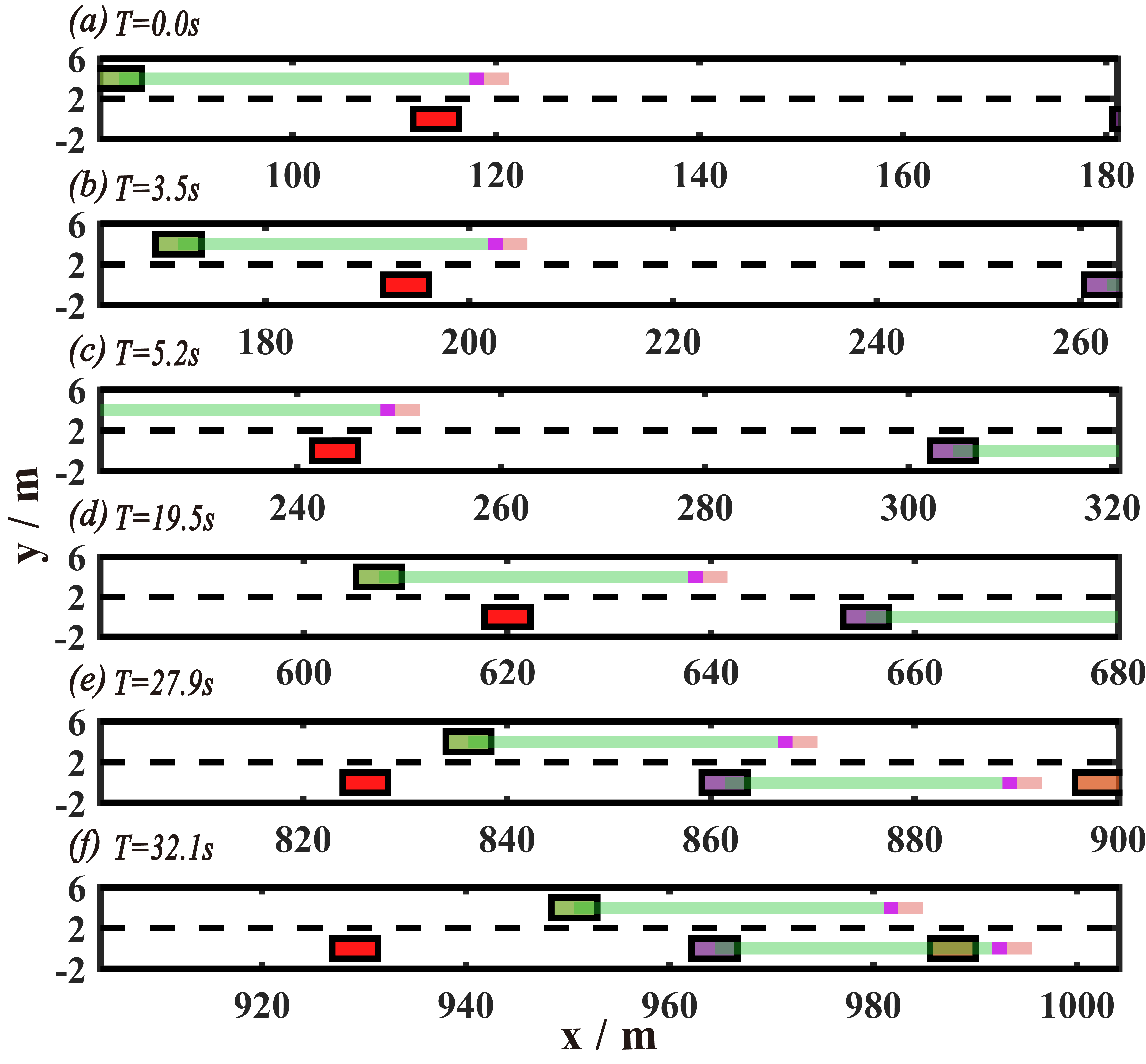}
        \caption{Spatial snapshots at representative instants (Case ~2).}
    \label{fig:case2_position}
\end{figure}

\begin{figure}[t]
    \centering
    \includegraphics[width=0.9\linewidth]{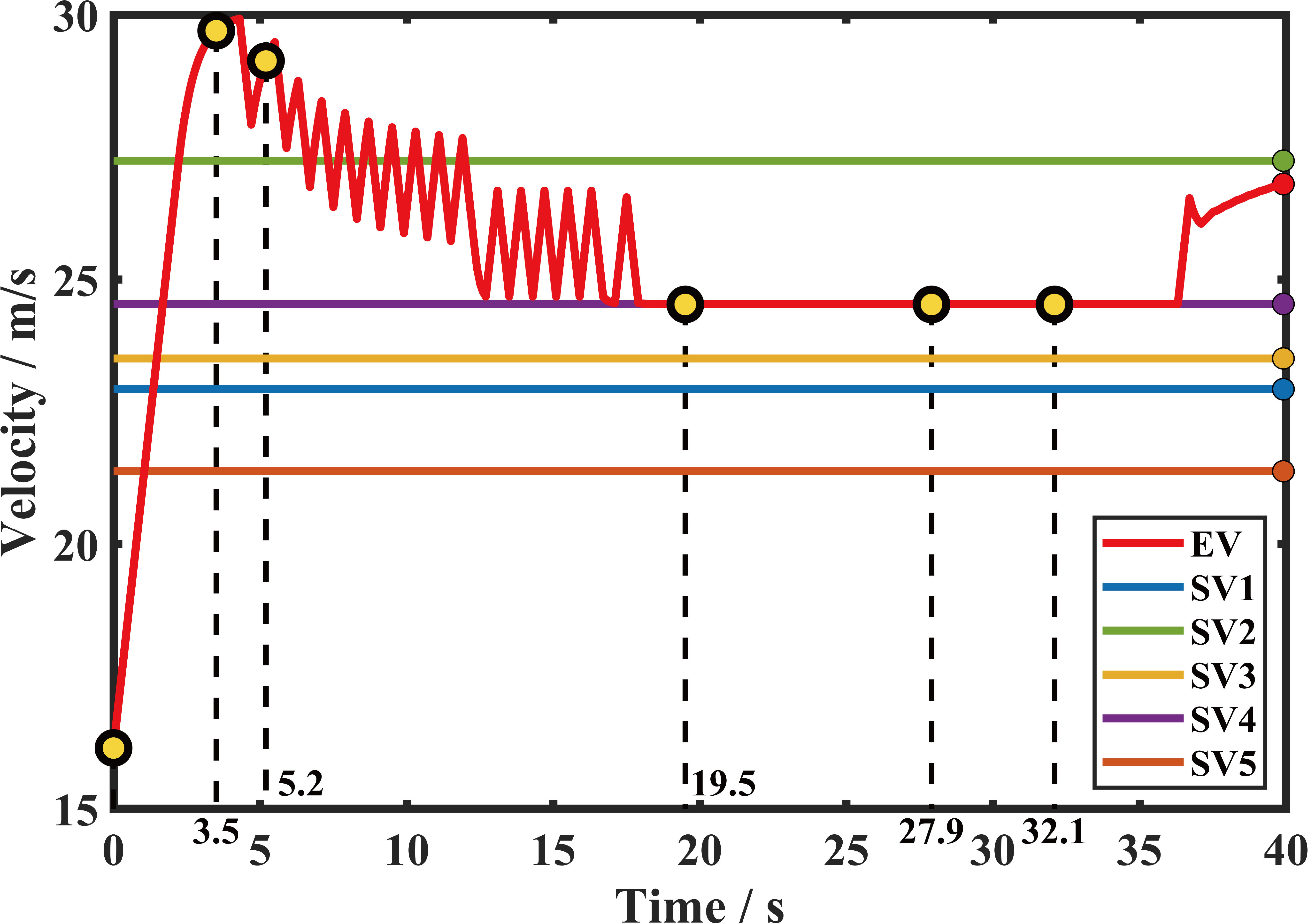}
    \caption{Velocity comparison (EV versus SVs) (Case 2). Yellow circles align with the snapshots in Fig.~\ref{fig:case2_position}.}
    \label{fig:case2_velocity}
    \vspace{-6pt}
\end{figure}

\begin{figure}[htbp]
    \centering
    \includegraphics[width=0.9\linewidth]{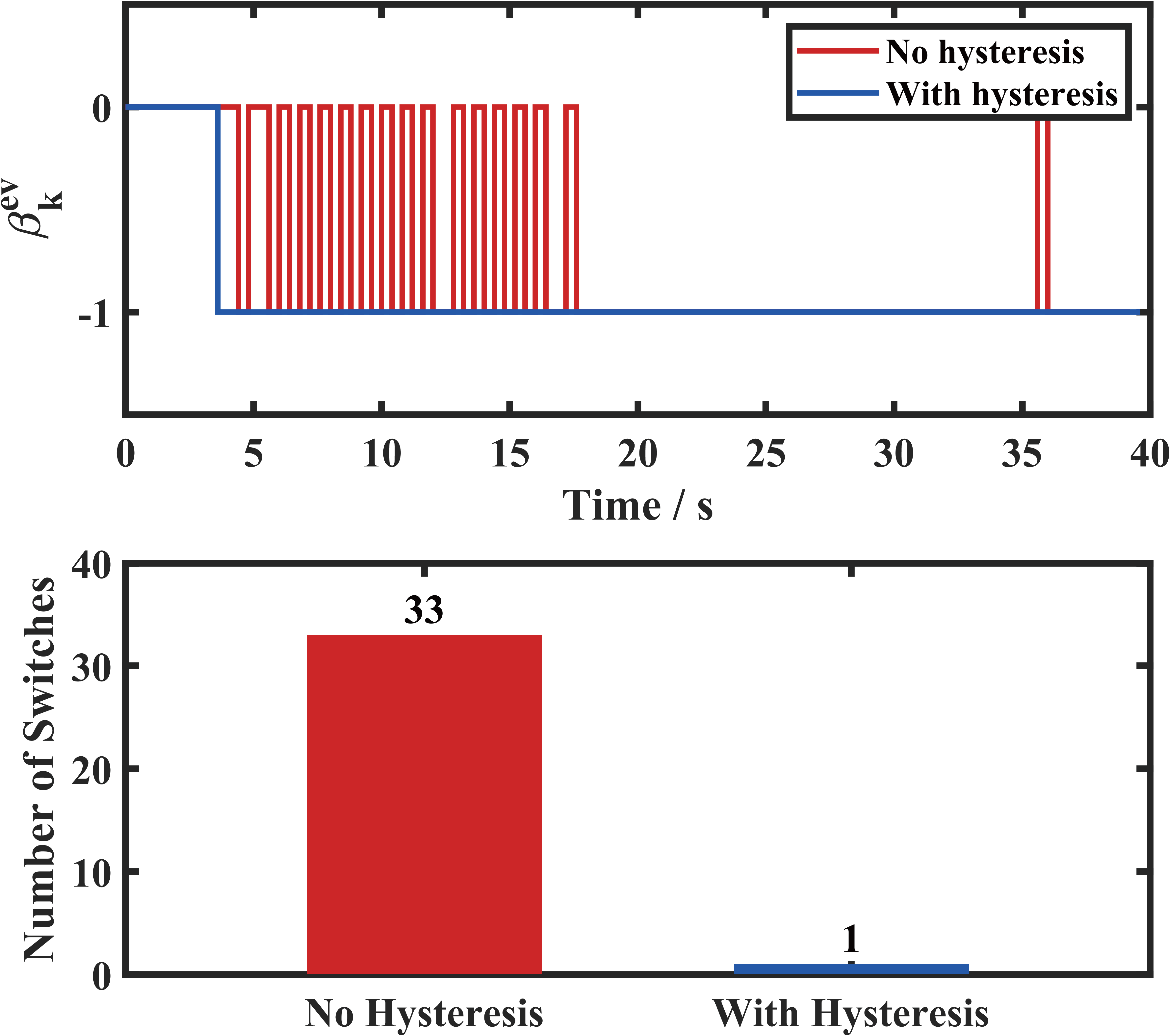}
    \caption{Ablation diagnostics: longitudinal maneuver state $\beta_k^{\text{\,ev}}$ over time for the no-hysteresis baseline versus\ the proposed hysteresis design.}
    \label{fig:ablation}
    \vspace{-6pt}
\end{figure}

We compare a no-hysteresis baseline with the proposed hysteresis design under identical initial conditions and identical objective weights.
As shown in Fig.~\ref{fig:case2_position} and Fig.~\ref{fig:case2_velocity}, 
removing hysteresis causes pronounced EV speed fluctuations after $t\!\approx\!3.5$\,s. Notably, at \(t\!\approx\!32.1\) s the hysteresis-enabled EV has already completed the lane change, while the no-hysteresis baseline hesitates. Fig.~\ref{fig:ablation} plots the longitudinal maneuver state: the baseline alternates rapidly between decelerate and cruising, while with hysteresis $\beta_k^{\text{\,ev}}$ holds steady and changes once decisively. Hysteresis reduces the number of longitudinal maneuver $\beta_k^{\text{\,ev}}$ switches from  $33$ to $1$, effectively eliminating throttle-brake switching.

\subsection{Case 3: Randomized Scenario Evaluation}
To stress-test the framework, we construct 18 scenario configurations by varying three factors: lane count $\{2,3,4\}$, number of SVs $\{5,8,10\}$, and velocity range $[10,20]$\,m/s (urban) versus $[25,40]$\,m/s (highway). 
For each configuration, at least 325 randomized trials are performed, with initial longitudinal positions randomly sampled from $[0,500]$\,m and vehicles assigned to lanes independently.

Four performance metrics are reported: collision rate, Normal Optimization Rate (NOR), Global Soft Relaxation Rate (GSRR), and Fallback Rate (FBR), which respectively denote the proportions of decision steps solved by nominal MPC, by global relaxation, and by the deterministic fallback.
Table~\ref{tab:random_test_results} summarizes the outcomes of  8{,}050 trials and 622{,}376 decisions: the overall collision rate is just 0.05 \%, while NOR reaches 98.77\%, with only 0.93\% GSRR and 0.29 \% FBR.  Hence, the framework solves nearly all cases in the nominal optimization, rarely invokes bounded relaxation, and only exceptionally falls back to emergency maneuvers, demonstrating strong robustness and adaptability under diverse traffic conditions. 
Besides, we evaluate the computational efficiency of the proposed framework over more than 10{,}000 decision steps. 
The average MILP solver time is 0.059~s (max 0.174~s), with all instances solved within the 0.4~s update period, 
suggesting suitability for real-time implementation in practical autonomous driving settings.

Together, these three studies demonstrate that the proposed framework achieves stable longitudinal regulation, efficient and stable maneuver execution, and strong robustness under diverse traffic conditions.
\begin{table*}[!t]
\centering
\captionsetup{font=footnotesize, skip=2pt}
\caption{Randomized Test Results Under Different Traffic Conditions (Case~3)}
\setlength{\tabcolsep}{5pt}
\label{tab:random_test_results}
\begin{threeparttable}
\begin{tabular}{cccccccccc}
\toprule
\textbf{ID} & \textbf{Lane Count} & \textbf{Velocity Range (m/s)} & \textbf{SV Count} & \textbf{$N$} & \textbf{Decision Steps} & \textbf{Collision Rate (\%)} & \textbf{NOR (\%)} & \textbf{GSRR (\%)} & \textbf{FBR}(\%) \\
\midrule
1  & 2 & [10, 20] & 5  & 500 & 37500  & 0.00\% & 99.84\% & 0.13\% & 0.03\% \\
2  & 2 & [10, 20] & 8  & 375 & 28125  & 0.00\% & 98.89\% & 0.87\% & 0.24\% \\
3  & 2 & [10, 20] & 10 & 325 & 24375  & 0.00\% & 98.04\% & 1.28\% & 0.67\% \\
4  & 2 & [25, 40] & 5  & 325 & 24375  & 0.00\% & 98.75\% & 1.05\% & 0.20\% \\ 
5  & 2 & [25, 40] & 8  & 600 & 44964  & 0.17\% & 96.20\% & 2.75\% & 1.06\% \\
6  & 2 & [25, 40] & 10 & 375 & 28094  & 0.27\% & 93.34\% & 4.39\% & 2.27\% \\
7  & 3 & [10, 20] & 5  & 500 & 37500  & 0.00\% & 99.97\% & 0.01\% & 0.02\% \\
8  & 3 & [10, 20] & 8  & 350 & 26202  & 0.29\% & 99.77\% & 0.15\% & 0.07\% \\
9  & 3 & [10, 20] & 10 & 325 & 24375  & 0.00\% & 99.59\% & 0.36\% & 0.05\% \\
10 & 3 & [25, 40] & 5  & 425 & 31875  & 0.00\% & 99.58\% & 0.38\% & 0.04\% \\
11 & 3 & [25, 40] & 8  & 625 & 46875  & 0.00\% & 98.76\% & 1.04\% & 0.20\% \\
12 & 3 & [25, 40] & 10 & 525 & 58125  & 0.00\% & 97.96\% & 1.67\% & 0.37\% \\
13 & 4 & [10, 20] & 5  & 500 & 37500  & 0.00\% & 99.99\% & 0.01\% & 0.00\% \\
14 & 4 & [10, 20] & 8  & 325 & 24375   & 0.00\% & 99.98\% & 0.02\% & 0.00\% \\
15 & 4 & [10, 20] & 10 & 375 & 28116  & 0.27\% & 99.95\% & 0.04\% & 0.01\% \\
16 & 4 & [25, 40] & 5  & 425 & 31875  & 0.00\% & 99.69\% & 0.26\% & 0.06\% \\
17 & 4 & [25, 40] & 8  & 600 & 45000  & 0.00\% & 99.46\% & 0.53\% & 0.01\% \\
18 & 4 & [25, 40] & 10 & 575 & 43125  & 0.00\% & 98.88\% & 1.01\% & 0.11\% \\
\midrule
\textbf{Overall} & -- & -- & -- & \textbf{8050} & \textbf{622376} & \textbf{0.05\%} & \textbf{98.77\%} & \textbf{0.93\%} & \textbf{0.29\%} \\
\bottomrule
\vspace{-18pt}
\end{tabular}
\end{threeparttable}
\end{table*}

\section{Conclusion}\label{Sec_con}

Unlike prior approaches that rely on ad hoc safety margins or isolated fixes, this paper establishes a unified decision-making framework in which HMDP-based modeling, MPC optimization, IDM-hysteresis safety logic, and a two-layer recovery scheme are co-designed. The effectiveness of this framework has been validated through diverse simulation studies, confirming robustness and continuity across heterogeneous traffic conditions.

A current limitation is the imperfect coordination between the high-level decision layer and the low-level control layer. In rare cases, the low-level controller fails to fully interpret the high-level maneuver commands and generate appropriate trajectories, resulting in occasional collisions (0.05\%). Future work will focus on refining the intermediate layer to better bridge high-level decisions and low-level execution, thereby enhancing the overall integrity of the framework.








\begin{thebibliography}{99}
\bibitem{chen2024review} S. Chen, X. Hu, J. Zhao, R. Wang, and M. Qiao, ``A review of decision-making and planning for autonomous vehicles in intersection environments,'' \textit{World Electric Vehicle Journal}, vol. 15, no. 3, p. 99, 2024.

\bibitem{wang2023decision} Y. Wang, J. Jiang, S. Li, R. Li, S. Xu, J. Wang, and K. Li, ``Decision-making driven by driver intelligence and environment reasoning for high-level autonomous vehicles: a survey,'' \textit{IEEE Transactions on Intelligent Transportation Systems}, vol. 24, no. 10, pp. 10362--10381, 2023.

\bibitem{shu2025decision} K. Shu, M. Ning, A. R. Alghooneh, S. Li, M. Pirani, and A. Khajepour, ``Decision making in urban traffic: A game theoretic approach for autonomous vehicles adhering to traffic rules,'' \textit{IEEE Transactions on Intelligent Transportation Systems}, 2025.

\bibitem{li2025investigation} T. Li, J. Ruan, and K. Zhang, ``The investigation of reinforcement learning-based end-to-end decision-making algorithms for autonomous driving on the road with consecutive sharp turns,'' \textit{Green Energy and Intelligent Transportation}, p. 100288, 2025.

\bibitem{cheng2024hierarchical} Z. Cheng, X. Zeng, H. Fang, G. Wang, and L. Dou, ``Hierarchical MPC-based motion planning for automated vehicles in parallel autonomy,'' \textit{Unmanned Systems}, vol. 12, no. 5, pp. 927--938, 2024.

\bibitem{wang2025high} X. F. Wang, J. Jiang, and W. H. Chen, ``High-level decision making in a hierarchical control framework: Integrating HMDP and MPC for autonomous systems,'' \textit{IEEE Transactions on Cybernetics}, 2025.

\bibitem{li2024integrated} S. Li, C. Liu, and W. H. Chen, ``An integrated MPC decision-making method based on MDP for autonomous driving in urban traffic,'' in \textit{Proceedings of the International Conference on Industrial Artificial Intelligence (IAI)}, pp. 1--6, 2024.

\bibitem{ammour2022mpc} M. Ammour, R. Orjuela, and M. Basset, ``A MPC combined decision making and trajectory planning for autonomous vehicle collision avoidance,'' \textit{IEEE Transactions on Intelligent Transportation Systems}, vol. 23, no. 12, pp. 24805--24817, 2022.

\bibitem{zhang2025mobil} X. Zhang, S. Zeinali, H. Wen, and G. Schildbach, ``MOBIL-based traffic prediction and interaction-aware model predictive control for autonomous highway driving,'' \textit{Control Engineering Practice}, vol. 164, p. 106434, 2025.

\bibitem{zhang2024interaction} X. Zhang, S. Zeinali, and G. Schildbach, ``Interaction-aware traffic prediction and scenario-based model predictive control for autonomous vehicles on highways,'' \textit{IEEE Transactions on Control Systems Technology}, 2024.

\bibitem{zhou2025robust} J. Zhou, Y. Gao, B. Olofsson, and E. Frisk, ``Robust motion planning for autonomous vehicles based on environment and uncertainty-aware reachability prediction,'' \textit{Control Engineering Practice}, vol. 160, p. 106319, 2025.

\bibitem{gilbert2021multi} A. Gilbert, D. Petrovic, J. E. Pickering, and K. Warwick, ``Multi-attribute decision making on mitigating a collision of an autonomous vehicle on motorways,'' \textit{Expert Systems with Applications}, vol. 171, p. 114581, 2021.

\bibitem{ames2019control} A. D. Ames, S. Coogan, M. Egerstedt, G. Notomista, K. Sreenath, and P. Tabuada, ``Control barrier functions: Theory and applications,'' in \textit{Proceedings of the European Control Conference (ECC)}, pp. 3420--3431, 2019.

\bibitem{alan2023control} A. Alan, A. J. Taylor, C. R. He, A. D. Ames, and G. Orosz, ``Control barrier functions and input-to-state safety with application to automated vehicles,'' \textit{IEEE Transactions on Control Systems Technology}, vol. 31, no. 6, pp. 2744--2759, 2023.

\bibitem{allamaa2024real} J. P. Allamaa, P. Patrinos, T. Ohtsuka, and T. D. Son, ``Real-time MPC with control barrier functions for autonomous driving using safety enhanced collocation,'' \textit{IFAC-PapersOnLine}, vol. 58, no. 18, pp. 392--399, 2024.

\bibitem{treiber2000congested} M. Treiber, A. Hennecke, and D. Helbing, ``Congested traffic states in empirical observations and microscopic simulations,'' \textit{Physical Review E}, vol. 62, no. 2, p. 1805, 2000.

\bibitem{liu2024comfort} S. Liu, X. Chen, Y. Zhao, H. Su, X. Zhou, and K. Zheng, ``Comfort-aware lane change planning with exit strategy for autonomous vehicle,'' \textit{IEEE Transactions on Knowledge and Data Engineering}, vol. 36, no. 7, pp. 2927--2941, 2024.

\bibitem{wei2010prediction} J. Wei, J. M. Dolan, and B. Litkouhi, ``A prediction- and cost function-based algorithm for robust autonomous freeway driving,'' in \textit{Proceedings of the IEEE Intelligent Vehicles Symposium (IV)}, pp. 512--517, 2010.

\bibitem{zheng2024highway} Z. Zheng, Y. Wang, and Y. Lin, ``Highway discretionary lane-change decision and control using model predictive control,'' in \textit{Proceedings of the Asian Control Conference (ASCC)}, pp. 195--201, 2024.

\bibitem{zhao2024real} W. Zhao, H. Wei, Q. Ai, N. Zheng, C. Lin, and Y. Zhang, ``Real-time model predictive control of path-following for autonomous vehicles towards model mismatch and uncertainty,'' \textit{Control Engineering Practice}, vol. 153, p. 106126, 2024.

\bibitem{samada2023robust} S. E. Samada, V. Puig, and F. Nejjari, ``Robust MPC-RG for an autonomous racing vehicle considering obstacles and the battery state of charge,'' \textit{Control Engineering Practice}, vol. 141, p. 105730, 2023.

\bibitem{geurts2023model} M. E. Geurts, A. Katriniok, E. Silvas, and W. P. M. H. Heemels, ``Model predictive control for lane merging automation with recursive feasibility guarantees,'' \textit{IFAC-PapersOnLine}, vol. 56, no. 2, pp. 4858--4864, 2023.

\bibitem{Calafiore2006DRCC}
G. C. Calafiore and L. E. Ghaoui, ``On distributionally robust chance-constrained linear programs,'' \textit{Journal of Optimization Theory and Applications}, vol. 130, no. 1, pp. 1--22, 2006.

\bibitem{Berkane2021Zeno} 
S. Berkane, A. Bisoffi, and D. V. Dimarogonas, ``Obstacle avoidance via hybrid feedback,'' \textit{IEEE Transactions on Automatic Control}, vol. 67, no. 1, pp. 512--519, 2021.





\end{thebibliography}
\end{document}